\documentstyle[12pt,psfig]{article}

\begin{document}

\setlength{\baselineskip}{17pt} 


\hfill NBI-97-44

\vspace{1cm}

\begin{center}
{\Large \bf Effects of collective expansion on light cluster
            spectra in relativistic heavy ion collisions}

\vspace{.5cm}

Alberto Polleri$^a$, Jakob P. Bondor$^a$ and Igor N. 
Mishustin$^{a,b}$.\\[.4cm] 
{\normalsize \it $^a$The Niels Bohr Institute, Blegdamsvej 17, DK-2100 
Copenhagen \O, Denmark.}\\[.05cm]
{\normalsize \it $^b$The Kurchatov Institute, Russian Scientific Center, 
Moscow 123182, Russia.}

\vspace{.5cm}

(Revised version, August 29, 1997)

\end{center}

\begin{abstract}
We discuss the interplay between collective flow and density profiles, 
describing light cluster production in heavy ion collisions at very
high energies. Calculations are performed within the coalescence model. 
We show how collective flow can explain some qualitative features of 
the measured deuteron spectra, provided a proper parametrization of the 
spatial dependence of the single particle phase space distribution is
chosen.
\end{abstract}

\vspace{.3cm}

\noindent PACS number(s): 25.75 -q, 25.75 -Ld

\vspace{.2cm}


\vspace{1cm}

In the course of a relativistic heavy ion collision a hot and 
dense fireball is created in the interaction region. Due to the high internal 
pressure it expands and cools down, finally disintegrating into hadrons.
The emergence of a collective flow can be considered as a signature that 
actually an extended piece of hot and dense matter is formed.
Among the products of the reactions a few light nuclei and antinuclei have
been observed(see, for instance, the Quark Matter '96 proceedings \cite{QM96}
and references therein), a very surprising fact for such high collision 
energies. A
common scenario employed to explain these observations is based on the 
coalescence model (for a review, see \cite{CK86}). A large amount of data on
composite particle spectra is accumulated now at intermediate collision 
energies \cite{R97}, where the multifragmentation of nuclei is a most
striking phenomenon. Although some trends in the fragment spectra are similar
to light cluster spectra in relativistic collisions, the mechanism of cluster 
production is in general very different in these two energy domains. In
particular, the coalescence picture does not really work for intermediate mass
fragments.

In this letter we examine the effect of collective expansion on the final 
spectra of clusters, assuming that they are produced via coalescence. We also 
underline how their measurements can give an additional information on 
the latest stages of a relativistic nuclear collision.

In the attempt of describing light cluster production in heavy ion
reactions at very high energies, one encounters a somewhat subtle problem.
Because these composite objects, typically $d$, $\overline d$, $t$ and
$^3\!He$, are very loosely bound, they can only be formed at the very late
stage of the reaction. This is because the system is then quite dilute and
interactions with the environment are therefore rare, preventing the formed 
clusters from breaking-up.
On the other hand, it is known that light nuclei cannot be formed by 
scattering nucleons in free space, even when the process is gentle
enough, simply as a consequence of energy-momentum conservation. The 
formation of a bound state requires the presence of a third body 
which carries away an amount of energy equal to the cluster binding energy.
It is  also clear that the system cannot be arbitrarily dilute and 
there must be a density around which the formation process is optimized.

In the present study we assume that the production process is governed by 
two distinct factors. The early stages of
the reaction and therefore the way particles are produced and emitted are
parametrized via a many-body phase space distribution. This is what we call
source funcion and it represents the probability that $A$ nucleons are emitted
at a given phase space point. The source function is taken at a sufficiently
late time, such that the conditions previously discussed are fulfilled.
The probability that these particles form a bound state, is taken as the
overlap between the cluster and the $A$-nucleon wave functions. This
framework has been quite commonly used and it is well described in \cite{LH94}.

Assuming that nucleons are emitted uncorrelated, one can factorize 
the generic $A$-body distribution as a product of single particle ones.
Denoting a phase space point and the corresponding measure as
\begin{equation}
x_i = (\vec{r}_i,\vec{p}_i)\ ,\ \ dx_i = 
\frac{d^3\vec{r}_i d^3\vec{p}_i}{(2 \pi)^3}\ ,\ \ i = 1, \dots, A\ ,
\end{equation}
we can write the phase space distribution of mass-$A$ clusters as
\begin{equation}
f_A(\vec{r},\vec{p}) = \int\! \prod_{i = 1}^A\,dx_i\,f(x_i)\ {\cal P}_A
(x_1, \dots, x_A; \vec{r},\vec{p})\ .
\end{equation}
This formula expresses the fact that, among all $A$-particle states, 
represented by $\prod_{i = 1}^A\,f(x_i) $, some can become bound with a 
probability ${\cal P}_A$. The integration goes over all phase space points,
where particles are emitted. The formation probability ${\cal P}_A$ is 
obtained by squaring a
corresponding quantum-mechanical amplitude, as done in \cite{M87,DHSZ91}. 
Below we adopt an approximation motivated by comparing the ranges of
variation in phase space of two factors, the single particle
distribution $f$ and the formation probability ${\cal P}_A$. Obviously the
first quantity has a much bigger range of variation than the second one,
especially when considering a large and 
hot system. This allows us to set $\vec{r}_i =  \vec{r}$ and $\vec{p}_i = 
\vec{p} / A$, for $i = 1,\ \dots,\ A$, and writing the general formula for 
the phase space distribution of mass-$A$ clusters in the form 
\cite{BJKG77}
\begin{equation}
f_A(\vec{r},\vec{p}) = \left[ f(\vec{r},\vec{p} /\! \mbox{\footnotesize $A$}) 
\right]^A\!\!.
\label{CLUSTER}
\end{equation}
This expression is the starting point of our subsequent analysis. 

Let us now specify the shape of the nucleon distribution in phase space. We 
assume that the system is in local thermal equilibrium, characterized by a 
temperature $T_0$, considered to be constant throughout the whole fireball
at the 
freeze-out stage of the reaction. We also assume that particles are subject 
to a collective velocity field, often also named collective flow. At very 
high energies it is generated by the partial transparency of nuclei,
along the longitudinal (beam) direction, and by the pressure created 
in the hot overlap zone, in the transverse direction. Since the dynamics in
these two directions is very different, we disregard possible correlations
and represent the collective velocity field as a sum of two independent
contributions,
\begin{equation}
\vec{v}(\vec{r}) = \vec{v}_L(\vec{r}_L) + \vec{v}_T(\vec{r}_T)\ ,
\label{SPLIT}
\end{equation}
where $\vec{r}_L = z\,\vec{e}_z$ and $\vec{r}_T = x\,\vec{e}_x + y\,\vec{e}_y$.
The nucleon momenta in a local frame $\vec{k}$ obey a thermal distribution
with temperature $T_0$. The transformation to a global frame is made with a
boost of velocity $\vec{v}(\vec{r})$. It is well known that the longitudinal
dynamics is highly relativistic, while the transverse expansion, in the first
approximation, can be considered non-relativistic, at least for nucleons.
Therefore, the nucleon momentum in the global frame can be written as the sum
of thermal and flow components as
$\vec{p}_T = \vec{k}_T + m\,\vec{v}_T(\vec{r}_T)$.
In the following discussion we ignore all issues related to the longitudinal
dynamics, focusing attention on the transverse direction. The transverse 
velocity field is parametrized as
\begin{equation}
\vec{v}_T(\vec{r}_T) = v_f \left(\frac{r_T}{R_0}\right)^{\alpha} \vec{e}_T
\label{FLOW}
\end{equation}
where $v_f$ and $R_0$ are the strength and scale parameters of flow and the
power-law
profile is characterized by the exponent $\alpha$. In building the phase space
distribution, we follow ref. \cite{CLZ96}. Assuming cylindrical symmetry we 
represent the nucleon density in a factorized form
\begin{equation}
\rho(\vec{r}) = N\,n_L(\vec{r}_L)\,n_T(\vec{r}_T)\ ,
\label{DENSA}
\end{equation}
such that $n_L$ and $n_T$ are normalized to $1$ in the respective domains. 
Using (\ref{CLUSTER}) we can now calculate the cluster phase space 
distribution function. Because of (\ref{SPLIT}) and (\ref{DENSA}), it also 
factorizes into longitudinal and transverse parts, namely
\begin{equation}
f_A(\vec{r},\vec{p}) = f_L(\vec{r}_L,\vec{p}_L)\,
f_T(\vec{r}_T,\vec{p}_T)\ .
\label{PHASE}
\end{equation}
The transverse contribution for clusters of mass number $A$ is therefore given
by
\begin{equation}
f_T(\vec{r}_T,\vec{p}_T) = 
(2\pi)^2\,\frac{\mbox{\small $1$}}{\mbox{\footnotesize $(2\pi M T_0)$}}\,
e^{- \mbox{\large $\frac{\left( \vec{p}_T - M\,\vec{v}_T \right)^2}
{2 M T_0}$}}\!\!\! n_A(\vec{r}_T)\ ,
\label{DIST}
\end{equation}
where $M = A\,m$ is the cluster mass and 
\begin{equation}
n_A(\vec{r}_T) = N_T(A)\,\left[ n_T(\vec{r}_T) \right]^A
\label{TRANDENS}
\end{equation}
is the transverse part of the cluster density, with the normalization factor
$N_T(A)$. Position and momentum of particles, completely uncorrelated in a 
purely thermal system, are now partially linked due to the presence of
collective flow. 

The transverse momentum spectrum of clusters is obtained by integrating
expression (\ref{PHASE}) over the whole volume and around a particular value 
${\overline p}_L$ of the longitudinal momentum\footnote{In a relativistic
formulation, a more familiar notation in terms of rapidity $y$ instead of
$p_L$ would appear, without anyway affecting our discussion on transverse
spectra.}.
The $p_T$-spectrum for clusters of mass number $A$ can be written in the form
\begin{equation}
\frac{d N_A}{dp_T^2} \mbox{\Large $|$}_{{\overline p}_L} = 
\nu_A({\overline p}_L)\,S_A(p_T)
\label{COMP}
\end{equation}
where $\nu_A({\overline p}_L)$ is the total number of clusters of mass $A$ 
produced at ${\overline p}_L$ and
\begin{equation}
S_A(p_T) = \int \! d^2\!\vec{r}_T
\ \frac{\mbox{\small $1$}}{\mbox{\footnotesize $(2\pi M T_0)$}}\,
e^{- \mbox{\large $\frac{\left( \vec{p}_T - M\,\vec{v}_T \right)^2}
{2 M T_0}$}}\!\!\! n_A(\vec{r}_T)
\label{SPECT}
\end{equation}
is the $p_T$-dependent part of the momentum spectrum. This last factor is
quite interesting. If flow were absent, the integral would give directly the 
Boltzmann factor, but the present case is, in general, more complicated, and 
a numerical treatment is needed. The most common parametrization used in the
literature combines a gaussian profile for the nucleon density \cite{M93} with
a linear profile ($\alpha = 1$) for collective flow (see \cite{CLZ96,CNH95},
especially
in relation with source parametrizations in interferometry studies). Only this 
choice allows for an analytical solution, which is the Boltzmann distribution
\begin{equation}
S_A(p_T) =
\frac{\mbox{\small $1$}}{\mbox{\footnotesize $(2\pi M T_{*})$}}\,
e^{- \mbox{\large $\frac{p_T^2}{2 M T_{*}}$}}\ ,
\label{BOLWRO}
\end{equation}
but now with the modified effective temperature (slope parameter)
\begin{equation}
T_{*} = T_0 + m\,v_f^2\ .
\label{WRONG}
\end{equation}
At first sight, this result looks appealing, but it actually contradicts both 
intuition and experiment. What is
wrong in the previous expression is the dependence of the slope $T_{*}$ only
on $m$ but not on $M$, as one would expect also by looking at the slopes 
extracted from measured spectra \cite{NA4497}. 
When performing the integral in (\ref{SPECT}), one notices an interesting 
feature. From (\ref{TRANDENS}) one sees that the density of clusters of mass 
$A$ is proportional to the $A$-th power of the nucleon density. In the case of 
a gaussian profile, one can see that the $A$-cluster density 
shrinks towards the central region.
This is easy to understand, since it is clearly more probable to make a
cluster where there are many particles than on the tail, where there are only
a few. Together with this, we choose a linear flow profile. This  is 
parametrized in (\ref{FLOW}) defining $v_f$ as the flow strength at the 
surface of the density distribution, characterized by the scale parameter 
$R_0$. What happens with the gaussian
profile is that the actual size of the cluster density has a smaller radius,
thereby picking up a smaller value for the flow velocity at the surface,
as compared to the case of single nucleons. This effect exactly cancels the 
A-dependence of $M$ in (\ref{WRONG}).
The other extreme would be to take a uniform density with a sharp 
surface at a given radius. Any power of this function would give the same 
profile, with the same radius. In other words it is equally probable to have
clusters everywhere in the region with non-zero density. As a consequence we 
expect in this case that the slope parameter will depend on $M$, since the 
flow velocity at the surface is the same for all clusters. This is not the
whole story. The flow profile could have a smaller exponent ($\alpha = 1/2$,
for example). Indications of such a behaviour have been observed in 
microscopic models of heavy ion collisions such as RQMD 
\cite{MJSSG95,MSSG97}.
Now some dependence of $T_{*}$ on $M$ would appear, even for a gaussian
density profile.

\begin{figure}
\centerline{\psfig{figure=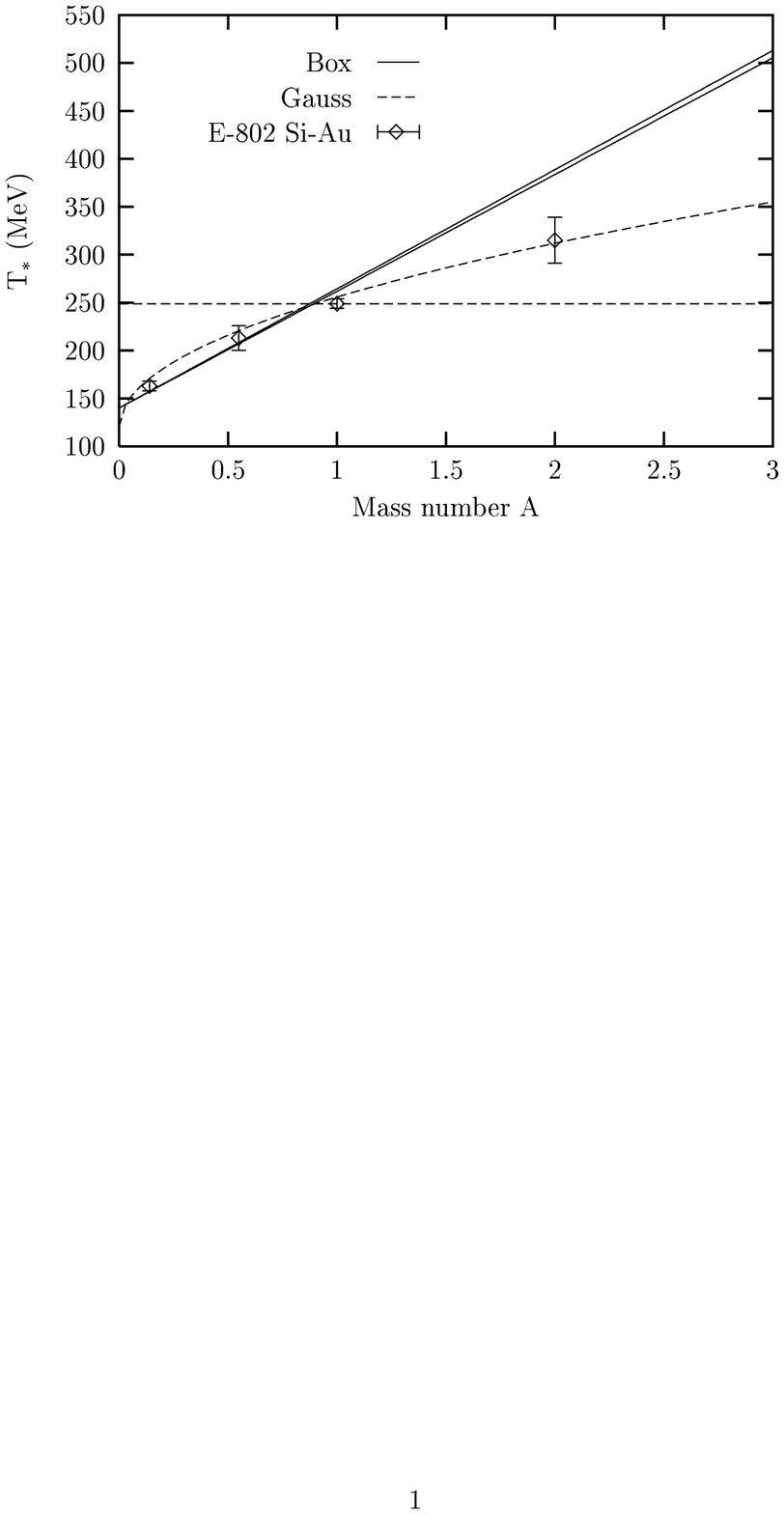,height=9cm}}
\protect\caption{Effective temperature $T_*$ as a function of mass number 
$A$. The top curve for each choice of density profile corresponds to 
$\alpha =$ 1/2, while the lower is for $\alpha =$ 1. The straight lines could
perfectly coincide if a more accurate choice of $v_f$ were made. This is
evident from (16), due to the $A$-independence 
of $\Lambda_{\alpha}^{[n]}$ for the box profile.}
\label{SLOP}
\end{figure}

Let us now look more closely at the interplay between flow and density 
profiles. It is clear that they cannot be considered independently because 
the density shape at a given time during expansion is the result of the 
particle motion characterized by the collective velocity field. The 
information about the profiles of density and collective velocity, can be 
extracted, in principle, from the energy spectra of different clusters 
\cite{R97}.
Unfortunately this is not an easy task because of the sensitivity of energy
spectra to all kinds of corrections \cite{SSH96,D9X}. We prefer a more global
analysis where the effective temperature (slope parameter) is extracted from
the $A$-dependence of the mean transverse energy\footnote{This approach was 
further developed in \cite{BFIM96}, including in the analysis also the
variance of the energy distribution.}
\begin{equation}
<E_T>_A\ = \int \frac{d^2 \vec{r}_T\,d^2 \vec{p}_T}{(2\pi)^2}
\ \frac{\vec{p}\,^2\!\!\!_T}{2M}\ f_T(\vec{r}_T,\vec{p}_T)\ .
\end{equation}
For a classical Boltzmann gas at temperature $T_0$ we have 
$<E_T>_A\ = T_0$, where the usual factor $3/2$ has changed to $2/2$ 
since we consider only the transverse degrees of freedom. In the present 
case we have instead
\begin{equation}
<E_T>_A \ =\ <E_T>_A^{th} + <E_T>_A^{fl}\ ,
\end{equation}
where the first term corresponds to the purely thermal, Boltzmann gas, 
while the second contribution arises due to the presence of flow (It vanishes
if we set $v_f = 0$). We define the effective temperature via
\begin{equation}
T_{*} =\ <E_T>_A \ = T_0 + \Lambda_{\alpha}^{[n]}(A)\,M\,v_f^2\ .
\label{EFFECT}
\end{equation}
The coefficient in front of the flow term depends on the flow parameter 
$\alpha$, is a functional of the transverse density and is a function of the
cluster mass number $A$. Different choices of density and flow profiles will 
result in a different $A$-dependence. The explicit expression for this 
coefficient is
\begin{equation}
\Lambda_{\alpha}^{[n]}(A) = \frac{1}{2} 
\int\! d^2 \vec{r}_T\,\left( \frac{r_T}{R_0} \right)^{2\alpha}\!\!
n_A(\vec{r}_T)\ .
\end{equation}
It can be calculated analytically for the two interesting cases of
gaussian and box profiles for the density, for all values of $\alpha$. In the 
first case one obtains
\begin{equation}
\Lambda_{\alpha}^{Gauss}(A) = \frac{2^{\alpha - 1}}{A^{\alpha}}\,
\Gamma(\alpha + 1)\ ,
\end{equation}
where $\Gamma$ is Euler's Gamma function.
One can readily see that for a linear flow profile ($\alpha = 1$) the 
coefficient is equal to $1/A$ and it exactly cancels the $A$ factor carried by 
$M$ in (\ref{EFFECT}). For lower powers of $\alpha$ the situation changes
one maintains a weak $A$-dependence. In the case of a box profile we obtain
\begin{equation}
\Lambda_{\alpha}^{Box}(A) = \frac{1}{2\alpha + 2}\ ,
\end{equation}
independent of $A$, as we expected after our previous discussion. Choosing
parameters according to Table \ref{TAB}, we illustrate the results in Figure
\ref{SLOP}, where the effective temperature is plotted as a function of mass
number $A$. The higher curves represent the extreme case of
a box-shaped density which gives the strongest $A$-dependence of $T_{*}$. The 
other extreme, as pointed out previously, is the gaussian density with linear
flow, which gives an $A$-independent $T_{*}$. Figure \ref{SLOP} suggests that 
the 
choice of a gaussian profile for the density and a flow profile with 
$\alpha = 1/2$ give the best agreement with the measured values of the slopes.

\begin{figure}
\centerline{\psfig{figure=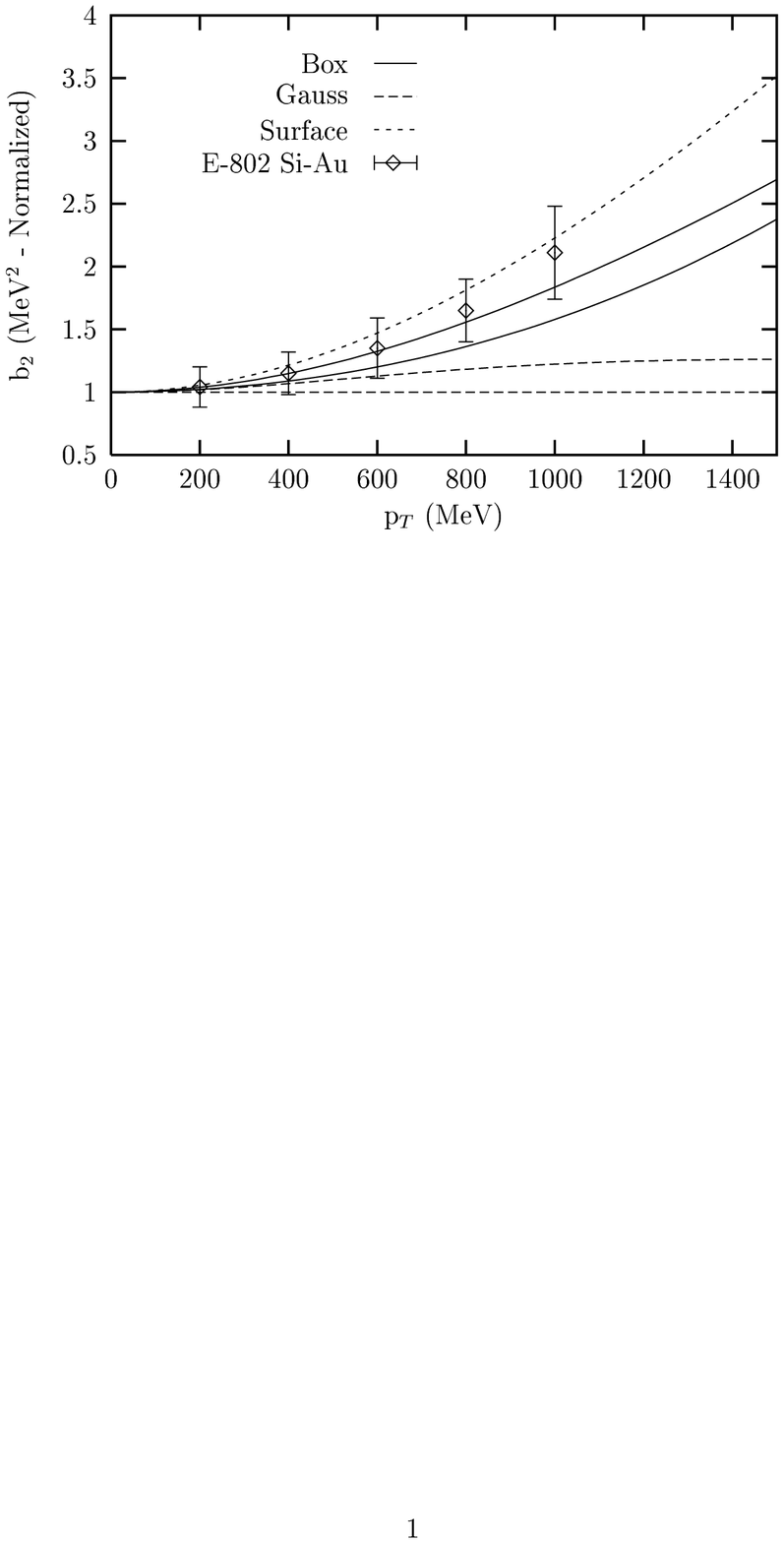,height=9cm}}
\protect\caption{$b_2$ factor as a function of transverse momentum. The top 
curve for each choice of density profile corresponds to 
$\alpha =$ 1/2, while the lower is for $\alpha =$ 1. The curve labelled
``Surface'' corresponds to integration over a spherical shell from $R_0 / 2$ to
$R_0$, in order to simulate surface emission.}
\label{PLOT}
\end{figure}

There is another interesting feature regarding cluster spectra, which
can be experimentally measured. In the early days 
of heavy ion physics the proportionality relation between cluster spectra and 
the corresponding powers of single particle spectra was quite well established 
\cite{CK86}. In recent experiments at much higher energies a momentum 
dependence in the proportionality constant $B_A$ was observed \cite{E80294}.
Namely, $B_A$ increases with increasing transverse momentum. In the present 
analysis we compare the transverse momentum spectrum of clusters of mass
number $A$ with the $A^{th}$ power of the single particle spectrum,
\begin{equation}
\frac{dN_A}{dp_T^2} \mbox{\Large $|$}_{{\overline p}_L} = 
b_A({\overline p}_L,p_T) 
\left( \frac{dN}{d(\frac{p_T}{A})^2} 
\mbox{\Large $|$}_{\frac{{\overline p}_L}{A}} 
\right)^A\!\!\!,
\end{equation}
where the first factor does not coincide with the usually quoted
$B_A$ because it is calculated for a small window around ${\overline p}_L$. 
We therefore discuss the $p_T$-dependence in this factor. Again, we 
perform the calculations with box and gaussian profiles, taking 
$\alpha = 1,\,1/2$. Using (\ref{COMP}) we obtain
\begin{equation}
b_A({\overline p}_L, p_T) = c_A({\overline p}_L)\,
\frac{S_A(p_T)}{\left[ S_A(p_T/A) \right]^A}\ ,
\end{equation}
where $c_A$ is a normalization factor which gives the order of magnitude of 
$b_A$, but does not affect its $p_T$-dependence. In Figure \ref{PLOT} we show
various plots of $b_2$, using the parameters from Table \ref{TAB}.
Although microscopic simulations are able to more or less reproduce this 
feature \cite{NKKSM96}, it is instructive to understand how it arises 
within the simple picture presented above. This behaviour is a pure
manifestation of collective flow, which only cluster measurements can reveal.
This effect depends on 
the relation between flow and density as we discussed above, resulting 
in turn in different shapes of momentum spectra for clusters and single 
nucleons. We emphasize again that the linear velocity profile and the 
gaussian shape for the density distribution are in contradiction with the 
$p_T$-dependence of $b_A$ (In this case both cluster and single particle 
spectra have the same slope). Also the choice $\alpha = 1/2$ does not help 
much, suggesting that a better understanding of the density shape is 
necessary. Therefore we indicate in Figure \ref{PLOT} that surface formation
of clusters, at a slightly earlier time with respect to the complete 
disintegration of the system, could improve ou scenario. This is done by
performing the spatial integration over a spherical shell from $R_0 / 2$ to
$R_0$ and is equivalent to having a density with a depleted central region,
as suggested in \cite{SNK95}. The actual situation is clearly a combination of
 this
early surface emission and final bulk disintegration and a consistent 
implementation of this aspect, together with a proper description of time 
evolution, is the subject of our current study.

In summary, we have shown that a suitable implementation of collective flow
can account for important qualitative features of light cluster spectra, 
measured in heavy ion collisions at very high energies, even though more has
to be done to build a consistent and quantitative description of the late
expansion stage. The observed
$A$-dependence of the slope parameters and the $p_T$-dependence of the
coalescence coefficients impose serious constraints on the spatial profiles
of the collective velocity and the particle density at the freeze-out stage.
The most common parametrizations for both flow and density profiles
fail to reproduce these features. Quantitative conclusions will be possible 
in the near future when cluster spectra for large and symmetric collision 
systems will be available.\\[.01cm]

We would like to thank Ian Bearden and the experimental heavy ion 
group at the Niels Bohr Institute for the pleasant collaboration. This work 
was supported in part by I.N.F.N. (Italy), the Carlsberg Foundation 
(Denmark) and the EU-INTAS grant 94-3405.

\newpage

\begin{table}
\protect\caption{Parameters used to perform the calculations. They are not
chosen in order to fit the data but only to show the qualitative beaviour in 
the following figures.}
\begin{center}
\begin{tabular}{|c|c|c|c|c|} \hline
Density   & $\alpha$ &   $v_f$  & $T_0$(MeV)  & $R_0$(fm) \\ \hline\hline
Box       &  \ 1/2\  & \ 0.63 \ &    140      &   8       \\ \cline{2-5}
          &  \  1 \  & \ 0.72 \ &    140      &   8       \\ \hline
Gauss     &  \ 1/2\  & \ 0.48 \ &    120      &   8       \\ \cline{2-5}
          &  \  1 \  & \ 0.34 \ &    140      &   8       \\ \hline
\end{tabular}
\end{center}
\label{TAB}
\end{table}

\end{document}